\documentstyle[aps,12pt,psfig]{revtex}
\begin{document}
\draft
\title{\large \large  \bf Saturation of product's exoticity in compound nuclear
reactions\\ and its role in the production of new $n$-deficient nuclei\\ 
with radioactive
projectiles.}
\vskip 1.3cm
\author{\bf Premomoy Ghosh, Alok Chakrabarti, V. Banerjee,
 Arup Bandyopadhyay,\\ D. N. Basu, Debasis Bhowmick}
\address{\bf \it Variable Energy Cyclotron Centre, \\ 1/AF Bidhan  Nagar, 
Calcutta - 700 064\\ India}
\maketitle
\vskip 1cm
\begin{abstract}
Representation in terms of a new parameter, exoticity, a measure of
$n$-deficiency or $p$-richness, clearly brings out
the saturation tendency of the product's maximum exoticity in a compound nuclear
reaction as the compound nucleus is made more and more exotic using radioactive
projectile (RIBs). The effect of this saturation on the production of new
proton-rich species with RIBs over a wide $Z$-range has been
discussed.
\end{abstract}

PACs: 25.60.Dz, 21.10.Dr


The compound nuclear reaction has been used extensively in the last two
decades for producing neutron-deficient ($n$-deficient) or proton-rich 
($p$-rich)
nuclei away from the $\beta$-stability. The fusion of two $\beta$-stable
heavy-ions leads in most cases to a neutron-deficient compound system and
the  lightest (most $n$-deficient) compound nucleus of any atomic number $Z$
can be reached through rather symmetric combination of target and projectile,
e.g. $^{50}Cr + ^{54}Fe \rightarrow ^{104}Sn$, $^{58}Ni + ^{74}Se
\rightarrow ^{132}Sm$, $^{64}Zn + ^{96}Ru \rightarrow ^{160}W$ etc.
The evaporation of a neutron from such a compound nucleus (CN)
 takes the residue or the
product towards the $p$-drip line while the evaporation of a proton or 
a $\alpha$-
particle brings it closer to the $\beta$-stability line as compared to the
said
compound nucleus. The Coulomb barriers (CB) for proton and $\alpha$ usually make
the evaporation of these particles energetically more costly compared to
neutron evaporation and, therefore, neutron evaporation is usually more
favoured. To the extent that the $n$-evaporation dominates one always
gains, vis-a-vis production of exotic species, by choosing appropriate
projectile-target combinations leading to the formation of lightest
compound systems.   
\vskip 0.3cm

        The possible availability of low energy (around Coulomb barrier)
radioactive ion beams (RIBs) in near future will certainly allow
formation of even lighter (as compared to the 'lightest' CN systems
possible with
stable projectile - stable target combinations)
CN systems and it is important to assess to what
extent these lighter CN systems can help in the production of new $n$-deficient
nuclei in or around the $p$-drip line. In other words it is important
to assess to what extent the naive expectation "more exotic the compound
nucleus, more exotic is the product" can be extrapolated as the compound
nucleus becomes more and more lighter.
\vskip 0.3cm

        The above expectation is known to hold for compound systems 
upto a certain distance
from the $\beta$-stability line but as one moves farther and farther
away the binding energy of the last proton decreases very rapidly and that
of the neutron increases sharply \cite{ref1}. For any given $Z$ if 
the mass number $A$ of 
the compound system is less than a certain value, the energy cost of 
the last proton (binding energy + CB for proton) becomes actually
lower than that of a neutron making thereby $p$-evaporation a
more likely process. The formation of compound systems beyond a certain
extent of $n$-deficiency, thus, may not lead, given any limit of
production cross-section, to the production of more $n$-deficient
products. Intuitively, therefore, there is a possibility of saturation in
the $n$-deficiency of the products. It is important to note that one needs to
set a lower limit for the production cross-section because to the extent
that no basic conservation laws (e.g. charge no., mass no. etc.) are
violated, products of any exoticity can be obtained, in principle, from
a given compound nucleus if one stops bothering whether the production
cross section is 1 mb, or say a millionth of a millibarn.

The lightest possible compound nuclei
with stable projectile - stable target combinations and obviously the
even lighter compound nuclei which can be formed with $p$-rich RIBs
fall mostly in the domain where effective separation energy of the last
proton is either equal to or less than that of the neutron. The product
pattern in such cases, therefore, are expected to show the effect of the
possible saturation, that is the formation of more $n$-deficient
compound systems may not lead, given any limit of production
cross-section, to more $n$-deficient products.
\vskip 0.3cm

        To see whether there is indeed any saturation or not one needs, 
at first,
to coin a definition of $n$-deficiency which is independent of $Z$. This is
because the products of a given compound nucleus can have a range of $Z$
values starting from $Z_{CN}$ ($Z$ of the CN ) to upto say, 7 or 8 units of 
atomic number less
than $Z_{CN}$. To compare which one of any two products of different $Z$
is more $n$-deficient we need a $Z$-independent description of $n$-deficiency
so that the products of different $Z$'s can be considered in the same 
footing. This can be achieved by defining a new parameter which we
prefer to call "exoticity".
\vskip 0.3cm

        Keeping in mind that the absolute value of the $n$-deficiency,
$A_{s} - A$ (where $A$ is the mass number of the nucleus of atomic number
$Z$ and $A_{s}$ is the mass number of the most $\beta$-stable isotope for the
same $Z$) alone can not be taken to be a measure of 'exoticity' of the
compound nucleus or the product since the $p$-drip line is only a few
neutrons away at lower $Z$ values while it is a few tens of neutrons away
at higher $Z$'s, we choose to define the 'exoticity' as,
\begin{eqnarray}
\zeta&=&1-(A-A_{d} )/(A_{s}-A_{d})\nonumber
\end{eqnarray}
where $A$ is the mass number of the nucleus of atomic number $Z$ and $A_d$
is the mass number of the isotope at the drip-line corresponding to  the
same $Z$. The exoticity is equal  to  1  on  the
drip-line, is zero on the stability  line and is greater than one beyond
the drip line. The mass  numbers $A_d$
were chosen from the compilation of Janecke and Masson \cite{ref2} which 
offers compilation of the proton drip-line
over a wide range.
\vskip 0.3cm

        The second hindrance in estimating the capability of a given compound
nucleus to produce exotic products with production cross-section greater
than any arbitrary chosen limit, is the excitation energy dependence of
the production cross-section and product distribution which are typical
to compound nuclear reactions. This makes any
description involving the products themselves not suitable for the
purpose (estimating the capacity of a CN to produce exotic products) since 
with increase in
the excitation energy more and more new channels are opened up changing
the products' distribution and also the cross-section of a given product.
\vskip 0.3cm

        To see whether or not any representation independent of excitation
energy is possible we have plotted in figure 1, as a typical example, 
the exoticity
of the maximum exotic product produced with cross-section greater than 1 mb
as a function of the excitation energy for different compound nuclei of
Ce (Cerium) having different exoticities. The cross-section values were
computed using the code ALICE \cite{ref3}. The plot reveals the
interesting feature that at lower CN exoticities an 
increase in the excitation energy leads to more exotic products 
but beyond a certain value
of CN-exoticity the products' exoticity becomes practically independent
of the excitation energy. It is important to note that the excitation
energy independence of the products' exoticity does not mean that the
'most exotic product' satisfying the minimum cross-section criterion
($\ge 1$ mb in this case) will remain the same at all the excitation
energies. At a given excitation energy there
will be one product (given the limit of production cross-section) which is 
most exotic. If one varies the excitation energy, the most exotic product
satisfying the minimum cross-section criterion may be a different isotope
but if one calculates its exoticity it will almost be the 
same as that of the most exotic product at the earlier excitation energy.
Further, the compound systems of Cerium for which the excitation 
energy dependence practically vanish or becomes very weak
are those which are lighter than the CN for which the separation energy of a 
neutron equals
to the effective separation energy of a proton, 
that is $B_{n} = {B_{p}}^{*}$
(${B_{p}}^{*} = B_{p} + CB$). For Cerium ${B_n}$ equals to ${B_{p}}^{*}$
for $\zeta_{CN} = 0.54$ ( $A$ = 127). Compound systems of other $Z$
values also exhibit similar dependence of products' exoticity on the
excitation energy.
\vskip 0.3cm

	In this study we attempt to estimate the production of very $p$-rich
exotic nuclei from compound nuclei formed by use of $p$-rich RIBs and 
having $Z$ in the range $50 \le Z_{CN} \le 82$. In this $Z$-range the Coulomb
barrier for protons is quite high (favouring production of exotic
species) and also the compound nuclear formation 
cross-section  and its subsequent decay by light particle emission
constitutes a major fraction of the total reaction cross-section for 
projectile energies not much above the Coulomb barrier.
\vskip 0.3cm 

	The compound nuclear systems of interest in the present study are those
which are more exotic than the compound nuclei
for which $B_{n}\sim {B_{p}}^{*}$. For example, for Ce the lightest CN
that can be reached with stable target - stable projectile combination
is $^{122}Ce$ for which $\zeta_{CN} = 0.75$. In the domain of our interest, 
therefore, an excitation energy independent description is possible if one
chooses a representation in terms of exoticity of the compound nucleus and
of the product rather than the usual representation in terms of $A$ and $Z$
of the products.
It is important to mention here that an accurate estimation of cross-sections 
of very exotic products is not possible no matter which one of the presently
available codes such as ALICE, CASCADE, PACE etc. is used for the
purpose. 
Our intention is,
therefore, not to predict the accurate cross-section values 
in a number of specific cases but to examine whether indeed there is any
saturation of products' exoticity and its implications on the production
of new nuclei with RIBs. 
\vskip 0.3cm

The dependence of $\zeta_p^{max}$ on
$\zeta_{CN}$ is shown in figure~2, where $\zeta$'s for odd-$Z$
products of maximum exoticity (that is $\zeta_p^{max}$) produced
from compound nuclei of representative even $Z$s and with cross-sections
greater than $1$~mb are plotted against corresponding $\zeta_{CN}$'s.
Exoticity of compound nucleus greater than one 
(i.e. beyond the drip-line) has also been considered. This is because
the $p$-decay life time, due to the existence of CB is expected to be 
longer than CN decay time upto a certain distance beyond the $p$-drip
line and one can attempt to form such compound systems so as to produce
the maximum $p$-rich products. To decide how far beyond $\zeta_{CN}=1$
one can go, 
the ground state $p$-decay life times, which are dependent on barrier
(Coloumb and centrifugal) penetration probabilities, for the compound
nuclei beyond the $p$-drip line have been calculated for various angular
momenta using WKB approximation. Only those compound
nuclei for which the ground state $p$-decay life times have been found
 (for $l=0$, to ensure a very conservative calculation) to be greater
than $10^{-14}$ sec. (once again to ensure a conservative estimate) 
are considered.
It can be seen from figure~2 that at each $Z_{CN}$, beyond a certain
$\zeta_{CN}$, $\zeta_p^{max}$ shows a saturation tendency and the value
of $\zeta_p^{max}$ at saturation increases with $Z$. The $\zeta_p^{max}$
at saturation, as expected, moves towards higher values as $Z$
increases. 
The gradient of the curves at various $Z_{CN}$'s are  
artefact of the relative binding energies or the evaporation
probabilities of mainly protons and alphas at corresponding $Z_{CN}$'s.
\vskip 0.3cm

	The curves shown in figure~2 clearly bring out the limitation,
as a result of the saturation, of the concept of forming more and more
exotic compound systems for the production of more and more exotic
$p$-rich species. 
The effect of the saturation is however not so serious
vis.a.vis production of odd $Z$ nuclei on or around the drip-line. For
odd $Z$ nuclei,  
the drip line can be reached for all nuclei having $Z \ge 51$ with
cross-sections $\ge$ 1mb. This is shown in fig.3 where odd $Z$ products
of maximum exoticity that can be produced with RIBs which are 4-neutron
away (deficient) from the lightest $\beta$-stable isotopes are shown 
in the $N$-$Z$
plane with two different cross-section limits of 1 mb and 0.01 mb. In
the cross-section limit of 0.01 mb the drip-line can be reached for all
nuclei with $Z \ge 45$ with only 4 neutron deficient RI projectiles.
\vskip 0.3cm

	For even $Z$ products, however, drip line can be reached with
the same projectiles (4-neutron deficient), as the calculation
reveals, only for $Z \ge 80$ with 1 mb cross-section limit. If the
cross-section limit is relaxed to $\ge$ 0.01 mb, the even $Z$ drip-line
can be reached for nuclei with $Z \ge 66$. The saturation thus affects seriously
the prospect of reaching the $p$-drip line with reasonable
cross-sections, say $\ge 10 \mu$b for all even $Z$ species with
$Z \le 66$.

	One can, however, consider the use of more exotic projectiles
to reach the even $Z$ drip line for $Z \le 66$, although the beam
intensity is likely to fall rather sharply with the exoticity. It
is important to note however that various other factors e.g. the signal
to noise ratio, 
detection efficiency, the type of measurement etc. together
decide the lower limit of cross-section and the intensity of RIB that
one needs in any given situation. The saturation, thus, in no way puts
any absolute restriction and should rather be considered as a hindrance
to be overcome by putting more efforts to increase beam intensity of 
RIBs, detection efficiency, background rejection etc.
\vskip 0.3cm

	It is important to note, in the context of discussions above,
that production of new nuclei with RI projectiles usually offer a number
of advantages as compared to the production of same nuclei with $\beta$-stable
projectiles. An estimation of these advantages is necessary to decide the
minimum usable beam intensity and the production cross-section ( the
product of these two represents a sort of quality factor) in a given
situation. To illustrate the possible advantages of RIBs we have plotted
in figure~4 the estimated cross-sections of isotopes of $Z=70$
products from three different compound systems of $W$, i.e, $Z=74$.
The compound nucleus of minimum exoticity $\zeta_{CN}$ = 0.76 is the
lightest one that can be 
formed from the stable projectile - stable target combination
($^{64}Zn + ^{96}Ru$). The other two compound nuclei of $\zeta_{CN}$'s
0.92 and 1.07 are formrd respectively from $^{60}Zn + ^{96}Ru$ and
$^{56}Zn + ^{96}Ru$. 
These curves clearly bring out 
the advantage of RIBs in terms of enhanced production cross-section
and in terms of enhancing the signal to background ratio. For example it can be
seen from  figure~4
that the production cross-section of $^{150}Yb$ (a  new  and  very
exotic nucleus) with RIB ($^{60}Zn$, $\zeta_{CN}$ = 0.93) 
is about $300$ times more than that with the  stable
projectile ($^{64}Zn$, $\zeta_{CN}$ = 0.76) and what  is  equally 
important for  experimental
measurements is the change in the relative production pattern.  With
stable projectile, the production cross-section of $^{150}Yb$ 
is almost  four  orders  of
magnitude  less  compared to the most favoured channel,
whereas in the RIB case ($\zeta_{CN} = 0.93$), $^{150}Yb$ is  produced
with the maximum cross-section. 
\vskip 0.3cm

Such situations are very  favourable in that they 
push down the lower limit of the needed RI beam intensity 
(for detection and other measurements) by several 
orders of  magnitude or conversely the lower limit of production 
cross-section is pushed down allowing measurements on even more exotic
species.
\vskip 0.3cm

        In this communication, we have attempted to address the question to what
extent one can hope to produce more exotic products by realising more and
more exotic compound systems using radioactive projectiles. It has been shown
that the exoticity of the product saturates beyond a certain value of the
exoticity of the CN. The value of the compound nucleus' exoticity at which
the saturation occurs depends on the atomic number $Z$ and moves, as expected,
towards higher value of exoticity as $Z$ increases. The conclusions reached
are practically independent of the excitation energy of the
compound nuclei as long as it is a few tens of MeV above the Coulomb barrier.
\vskip 0.3cm

        This new revealation has important consequences in the production of
new $p$-rich species using $p$-rich projectiles (RIBs). It
tends to limit, to an extent, the utility of very $p$-rich projectiles
vis-a-vis production of new $p$-rich species in or around the $p$-drip line.
While the saturation does not affect that adversely the production of new
odd $Z$ nuclei where the proton drip line can be reached with reasonable
cross-sections ($\ge 10 \mu$b) for all elements with $Z \ge 45$ with
only moderately $p$-rich projectiles (4 neutron deficient),
it does make production of $p$-drip line nuclei for even
$Z$ (especially for $Z\le 65$) difficult with 
moderate production cross-sections unless one uses quite exotic
projectiles.
\vskip 0.3cm

{\bf Acknowledgement:}
\vskip 0.1cm
 The authors gratefully acknowledge the support and
encouragement received from Dr. Bikash Sinha for this work. They are
also grateful to Dr. J.N.De and Dr. Santanu Pal for many helpful 
discussions and suggestions.

\newpage
{\bf Figure Captions.}
\vskip 0.3cm
	Figure-1. Dependence of maximum exoticity of products against the
excitation energy (in MeV) of compound nuclei of different $\zeta_{CN}$.

	Figure-2. The variation of maximum exoticity $\zeta_p^{max}$ for odd
 $Z$ products with the exoticity of even $Z$ compound nuclei 
 $\zeta_{CN}$.

	Fifure-3. The $\beta$-stability line along with the proton drip line and
lines showing the extent of production of exotic nuclei with RIBs which
are only four
neutron deficient compared to the lightest stable projectiles of
corresponding $Z$ (within the cross-section limits of 1mb and 0.01
mb).

	Figure-4. The cross-section of $Z=70$ isotopes against mass number
$A$ at representative $\zeta_{CN}$ values.
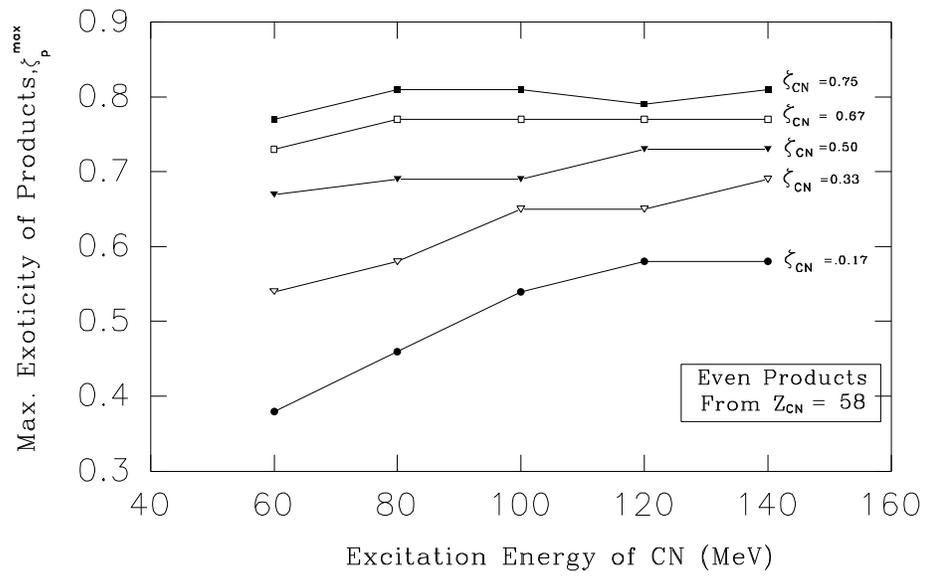
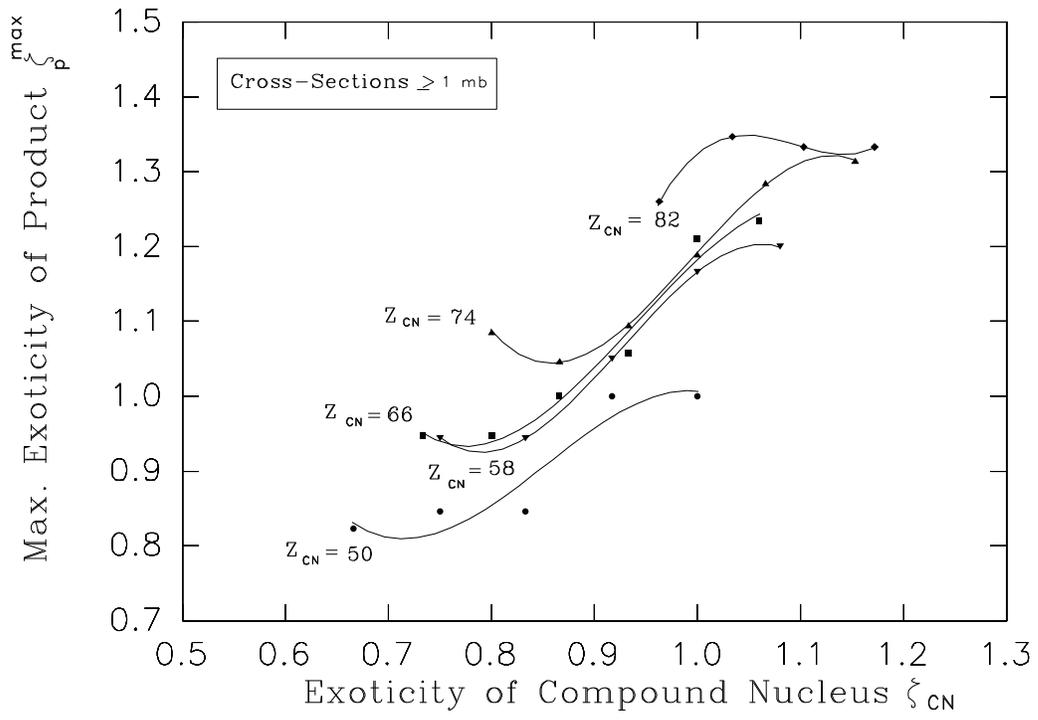
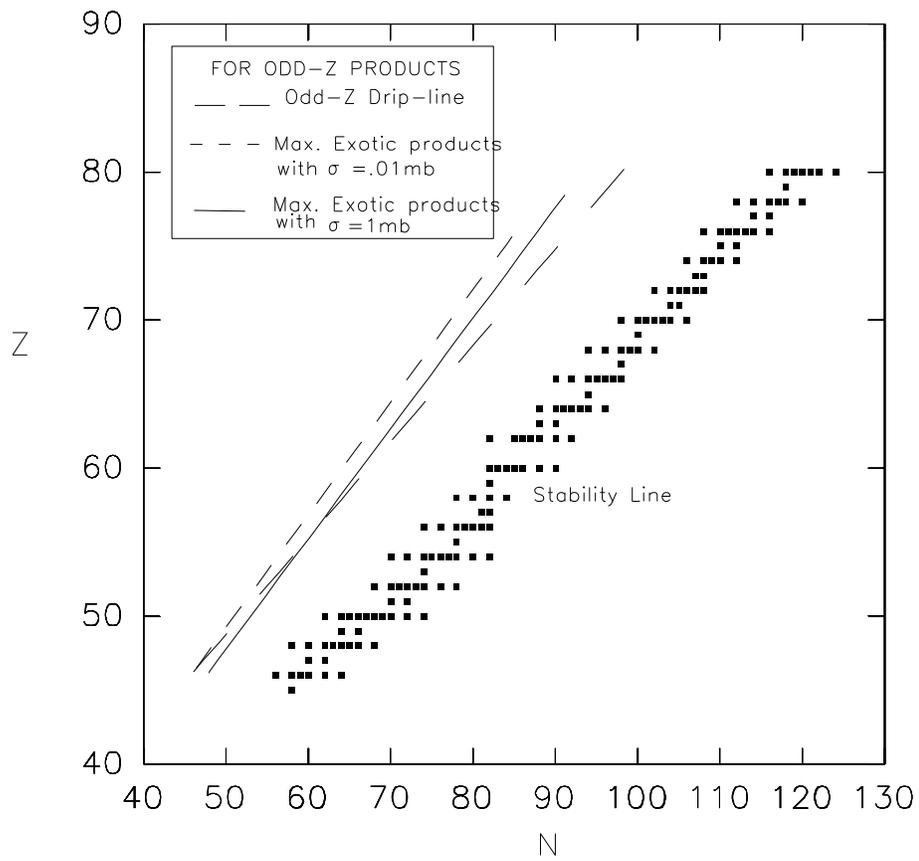
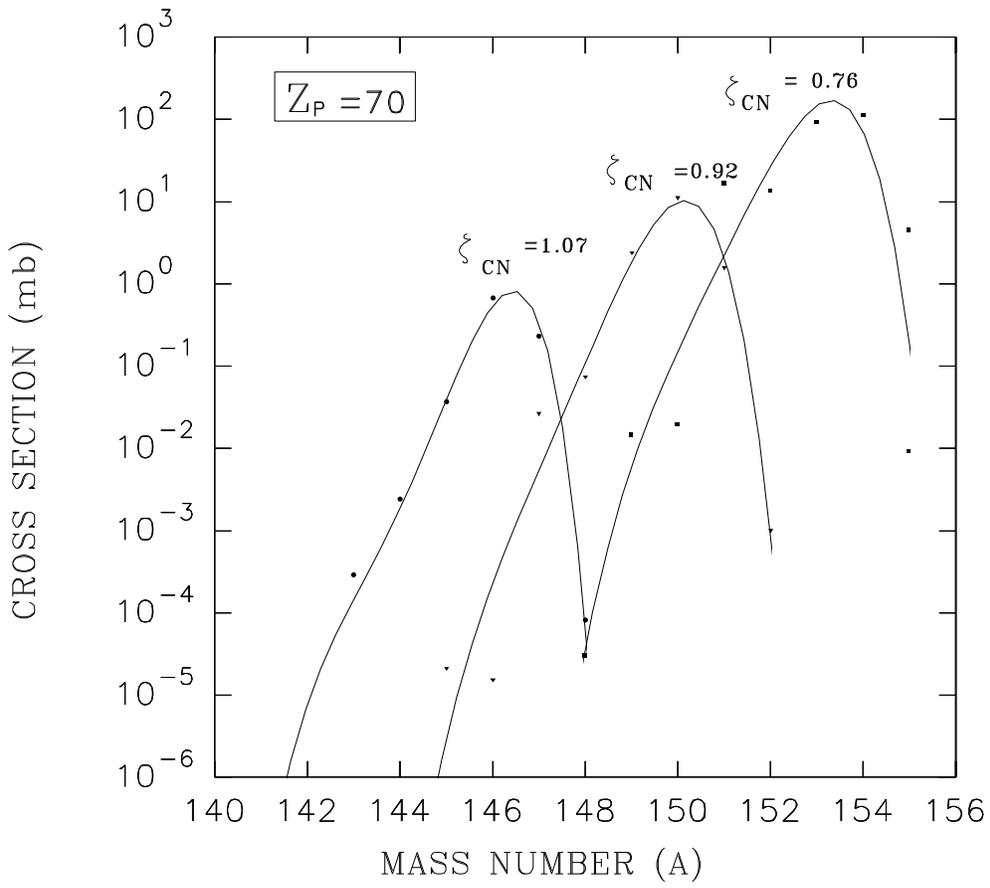
\end{document}